\documentclass[12pt, draftcls, onecolumn]{IEEEtran}
\ifCLASSINFOpdf
  \usepackage[pdftex]{graphicx}
  \DeclareGraphicsExtensions{.pdf,.jpeg,.png}
\else
  \usepackage[dvips]{graphicx}
\fi
\usepackage{subfigure}
\usepackage{epstopdf}
\usepackage[cmex10]{amsmath}
\usepackage{amssymb}
\usepackage{wasysym}
\usepackage{algorithm}
\usepackage{algorithmic}
\usepackage{array}
\usepackage{cite}
\usepackage{color}
\usepackage{url}
\usepackage{subfigure}
\usepackage{epsfig,latexsym}
\usepackage{flushend}
\usepackage{hyperref}
\usepackage{diagbox} 
\usepackage{booktabs}
\usepackage{etoolbox}

\begin{document}
%
\title{Beam Training based on Dynamic Hierarchical Codebook for Millimeter Wave Massive MIMO}

\author{Kangjian~Chen,~\IEEEmembership{Student~Member,~IEEE}, and  Chenhao~Qi,~\IEEEmembership{Senior~Member,~IEEE}\thanks{This work is supported in part by National Natural Science Foundation of China under Grant 61871119 and Natural Science Foundation of Jiangsu Province under Grant BK20161428. (\textit{Corresponding author: Chenhao~Qi})}
\thanks{Kangjian~Chen and Chenhao~Qi are with the School of Information Science and Engineering, Southeast University, Nanjing 210096, China (Email: qch@seu.edu.cn).} }

\markboth{}
{Shell \MakeLowercase{\textit{et al.}}: Bare Demo of IEEEtran.cls for Journals}

\maketitle

\begin{abstract}
Beam training based on hierarchical codebook for millimeter wave (mmWave) massive MIMO is investigated. Unlike the existing work using the same hierarchical codebook to estimate different multi-path components (MPCs), dynamic hierarchical codebooks which are updated according to the estimated MPCs are adopted. Firstly, a generalized hierarchical codebook design method is proposed. Then based on this method, a beam training method which dynamically updates the hierarchical codebook by removing the contribution of the estimated MPCs from the codebook is proposed. Simulation results verify the effectiveness of our method and show that the proposed method outperforms the existing ones in terms of the success detection rate of beam training.

\end{abstract}

\section{Introduction}
Millimeter wave (mmWave) massive MIMO considered as a promising technology for future wireless communications has drawn global attention due to its wide band and high transmission rate~\cite{CL1,ShuFengTWC}. At high frequency, mmWave signal experiences large path loss. However, it can be compensated by a massive MIMO antenna array which achieves directional beamforming and transmission. Note that the antenna array can be installed in a small area, since the higher frequency of mmWave signal leads to smaller wavelength and shorter antenna interval.

With the large dimension of antenna array for mmWave massive MIMO, conventional full digital precoding structure is difficult to achieve due to the hardware constraint from the expensive radio frequency (RF) chains. To reduce the number of RF chains, a hybrid precoding structure is widely adopted, where the large antenna array is driven by a small number of RF chains~\cite{Sparse2014}. Based on the hybrid precoding structure, two methods including compressed sensing (CS) based channel estimation and hierarchical codebook based beam training are typically used to acquire the channel state information (CSI) of mmWave massive MIMO~\cite{Sparse2014,Xiao2016Hierarchical,ZhenyuXiao2018}. The CS based channel estimation usually exploits the channel sparsity by sending training sequences and employs sparse recovery algorithms for sparse channel estimation. The hierarchical codebook based beam training is performed aiming at finding the best pair of transmitting beam and receiving beam based on a predefined hierarchical codebook, where a low-resolution codeword in the upper layer of the codebook covers several high-resolution codewords in the lower layer. Although the CS based channel estimation is more flexible when it is incorporated with sparse signal processing technique, the hierarchical codebook based beam training has larger receiving signal gain especially when the physical beam matches the channel path.

However, using the hierarchical codebook based beam training to estimate the mmWave massive MIMO channel with multi-path components (MPCs) is not trivial. To be specific, the contribution of previously estimated MPCs needs to be clearly removed before estimating new MPCs, which means that precise information of previously estimated MPCs is needed~\cite{Xiao2016Hierarchical,ZhenyuXiao2018}. Among the existing work, an adaptive CS (ACS) method is proposed for mmWave massive MIMO system, where the number of channel measurements can be reduced by designing a multi-resolution hierarchical codebook~\cite{Sparse2014}. In~\cite{Xiao2016Hierarchical}, a hierarchical search (HS) based beam training is proposed to acquire the CSI, which shows that HS performs much faster than ACS. To further improve the estimation precision, a multi-path decomposition and recovery (MDR) method which combines the advantage of HS and CS is proposed in~\cite{ZhenyuXiao2018}.

In this letter, we propose a beam training method based on dynamic hierarchical codebook for mmWave massive MIMO. Unlike the existing work using the same hierarchical codebook to estimate different MPCs, we use dynamic hierarchical codebook which is updated according to the estimated MPCs. To summarize, our contribution mainly includes: 1) A generalized hierarchical codebook design method is firstly proposed. 2) Then based on this method, we propose a beam training method which dynamically updates the hierarchical codebook by removing the contribution of the estimated MPCs from the codebook.

The notations are defined as follows. Symbols for matrices (upper case) and vectors (lower case) are in boldface. $\left[ \boldsymbol{a} \right] _n$ denotes the $n$th entry of a vector $\boldsymbol{a}$. $(\cdot)^{*}$, $(\cdot)^T$, $(\cdot)^H $, $|\cdot |$, $\|\cdot \|_2$, $\mathbb{C}$, $\mathbb{Z}$, $\mathbb{E}\{ \cdot \}$, $\backslash$ and $\mathcal{C}\mathcal{N}$ denote the conjugate, transpose, conjugate transpose (Hermitian), absolute value, $\ell_2$-norm, set of complex number, set of integer number, operation of expectation, operation of set exclusion and complex Gaussian distribution, respectively.

\section{System Model}
We consider a point-to-point mmWave massive MIMO system with $N_{\rm t}$ and $N_{\rm r}$ antennas at the transmitter and receiver, respectively. The antennas at both sides are placed in uniform linear arrays with half wavelength intervals.  To simplify the analysis, we assume both $N_{\rm t}$ and $N_{\rm r}$ in a power of two. With hybrid precoding and combining structure, the same $N_{\rm RF}$ RF chains at both sides are fully connected to the antenna array via phase shifters. In this work, we mainly consider the beam training in mmWave communications, where a training symbol $x$ with normalized power, i.e., $\mathbb{E}\{xx^{*}\}=1$, is transmitted. Then we express the  received signal as
\begin{equation}\label{system model}
y=\sqrt{P}\boldsymbol{w}_{\rm BB}^H\boldsymbol{W}_{\rm RF}^H\boldsymbol{H} \boldsymbol{F}_{\rm RF}\boldsymbol{f}_{\rm BB}x + \boldsymbol{w}_{\rm BB}^H\boldsymbol{W}_{\rm RF}^H \boldsymbol{\eta}
\end{equation}
where $\boldsymbol{f}_{\rm BB}\in \mathbb{C}^{N_{\rm RF}}$, $\boldsymbol{F}_{\rm RF}\in \mathbb{C}^{N_{\rm t}\times N_{\rm RF}}$, $\boldsymbol{w}_{\rm BB}\in \mathbb{C}^{N_{\rm RF}}$, $\boldsymbol{W}_{\rm RF}\in \mathbb{C}^{N_{\rm r}\times N_{\rm RF}}$, $\boldsymbol{H}\in \mathbb{C}^{N_{\rm r}\times N_{\rm t}}$ and $\boldsymbol{\eta}\in \mathbb{C}^{N_{\rm r}}$ denote the digital precoder, analog precoder, digital combiner, analog combiner, mmWave MIMO channel and additive white Gaussian noise (AWGN) vector with $\boldsymbol{\eta}\sim\mathcal{C}\mathcal{N}\left( \boldsymbol{0},\sigma _{\eta}^{2}\boldsymbol{I}_{N_{\rm r}} \right)$. The total power of the transmitter is $P$. It is common to suppose that both the precoder and combiner do not provide power gain, i.e., $\|\boldsymbol{F}_{\rm RF}\boldsymbol{f}_{\rm BB}\|_2=1$ and $\|\boldsymbol{W}_{\rm RF}\boldsymbol{w}_{\rm BB}\|_2=1$. The mmWave MIMO channel matrix is modeled as\cite{heath2016overview}
\begin{equation}\label{channel model}
\boldsymbol{H}=\sqrt{\frac{N_{\rm t} N_{\rm r}}{L}}
\sum_{l=1}^L \lambda_l\alpha(N_{\rm r},\Omega_l^{\rm r})\alpha(N_{\rm t},\Omega_l^{\rm t})^H
\end{equation}
where $L$, $\lambda_l$, $\Omega _{l}^{\rm r}$ and $\Omega _{l}^{\rm t}$ denote the number of MPCs, channel gain, channel angle-of-arrival (AoA), channel angle-of-departure (AoD) of the $l$th path, respectively. Define the physical AoD and AoA of the $l$th path as $\omega _{l}^{\rm t}$ and $\omega_{l}^{\rm r}$ respectively. We have $\Omega _{l}^{\rm t}=\cos \left( \omega _{l}^{\rm t} \right)$ and $ \Omega_{l}^{\rm r}=\cos \left( \omega _{l}^{\rm r} \right)$, regarding the half wavelength interval of adjacent antennas. Therefore, we have $\Omega_{l}^{\rm t}\in[-1,1]$ and $\Omega_{l}^{\rm r}\in[-1,1]$. The channel steering vector is defined as a function of $N$ and $\Omega$ as
\begin{equation}\label{steering vector}
\alpha\left( N,\Omega \right) =\frac{1}{\sqrt{N}}\left[ 1,e^{j\pi \Omega},\cdots ,e^{j\left( N-1 \right) \pi \Omega} \right] ^T
\end{equation}
where $N$ is the number of antennas and $\Omega$ is the channel AoD or AoA.

\begin{figure}[!t]
  \centering
  \includegraphics[width=100mm]{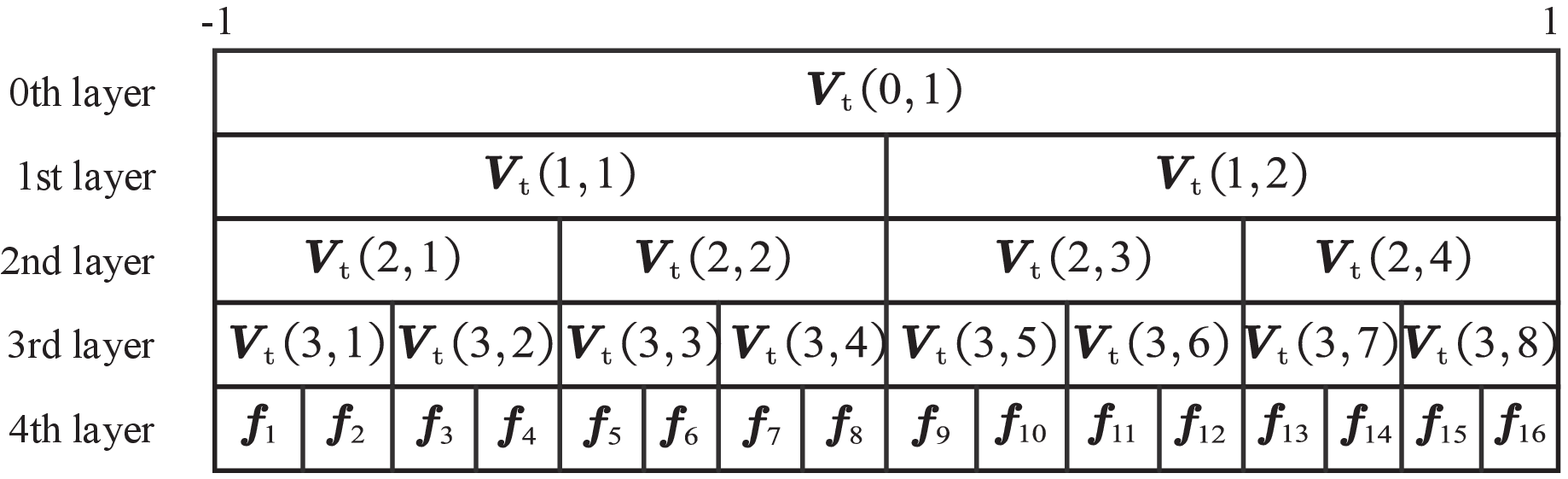}
  \caption{Illustration of a hierarchical codebook with $N_{\rm t}=16$.}
  \label{fig:HierarchicalCodebook}
\end{figure}

To fast acquire CSI of mmWave massive MIMO systems, hierarchical search based on hierarchical codebook is widely adopted. Hierarchical codebooks employed at the transmitter and the receiver are denoted by $\boldsymbol{V}_{\rm t}$ and $\boldsymbol{V}_{\rm r}$, respectively. As shown in Fig.~\ref{fig:HierarchicalCodebook}, a hierarchical codebook $\boldsymbol{V}_{\rm t}$ with $N_{\rm t}=16$ is illustrated, where $\boldsymbol{V}_{\rm t}(s,m)$ denotes the $m(m=1,2,\cdots2^s)$th codeword at the $s(s=0,1,\cdots,S)$th layer of $\boldsymbol{V}_{\rm t}$. In general, $S$ is the number of  layers of $\boldsymbol{V}_{\rm t}$ and can be determined by $S=\log_2{N_{\rm t}}$.  During beam training, the signal $x$ is transmitted using a codeword $\boldsymbol{v}$ from $\boldsymbol{V}_{\rm t}$. The received signal using a codeword $\boldsymbol{w}$ from $\boldsymbol{V}_{\rm r}$ can be expressed as
\begin{equation}\label{system model2}
y=\sqrt{P}\boldsymbol{w}^H\boldsymbol{H} \boldsymbol{v}x + \boldsymbol{w}^H\boldsymbol{\eta}
\end{equation}
where $\boldsymbol{v} \triangleq \boldsymbol{F}_{\rm RF} \boldsymbol{f}_{\rm BB}$ and $\boldsymbol{w} \triangleq \boldsymbol{W}_{\rm RF} \boldsymbol{w}_{\rm BB}$. To estimate $L$ MPCs, it is common to estimate them one by one using beam training based on hierarchical codebook~\cite{Sparse2014,Xiao2016Hierarchical,ZhenyuXiao2018}. Suppose $L_{\rm f}(L_{\rm f}<L)$ MPCs have already been estimated.  There are $L-L_{\rm f}$ MPCs to be estimated. Then \eqref{system model2} can be rewritten as
\begin{align}\label{system model3}
\setlength{\abovedisplayskip}{3pt} 
\setlength{\belowdisplayskip}{3pt}
y=\underset{\rm to~be~estimated}{\underbrace{\sqrt{P}\boldsymbol{w}^H\sum_{l=L_{\rm f}+1}^{L}\lambda_l\alpha(N_{\rm r},\Omega_l^{\rm r})\alpha(N_{\rm t},\Omega_l^{\rm t})^H \boldsymbol{v}x}}+\boldsymbol{w}^H\boldsymbol{\eta} \nonumber\\
+\underset{\rm estimated}{\underbrace{\sqrt{P}\boldsymbol{w}^H\sum_{l=1}^{L_{\rm f}}\lambda_l\alpha(N_{\rm r},\Omega_l^{\rm r})\alpha(N_{\rm t},\Omega_l^{\rm t})^H \boldsymbol{v}x}}.
\end{align}
Before estimating the $(L_{\rm f}+1)$th path, the contribution of the $L_{\rm f}$ estimated MPCs, i.e., ``estimated'' part in \eqref{system model3}, should be removed. The existing work usually computes the contribution of the estimated MPCs and then subtracts the ``estimated'' part from $y$ in \eqref{system model3}. However, it relies on the precision of the ``estimated'' part~\cite{Xiao2016Hierarchical,ZhenyuXiao2018}. Due to the accumulated estimation errors of channel AoA and AoD, the accurate estimation of $\lambda_l$ will be quite difficult. Unlike the existing work, in this letter, we avoid the substraction by dynamically updating the hierarchial codebooks according to the ``estimated'' part, which can effectively avoid the estimation of channel gain. 

\section{Multi-path Channel Beam Training based on Dynamic Hierarchical Codebook}
For the multi-path channel, we propose a beam training method based on dynamic hierarchical codebook. In the first subsection, we present a generalized hierarchical codebook design method, which is then used to design dynamic hierarchical codebook for beam training in the second subsection.

\subsection{Generalized Hierarchical Codebook Design}
Since the hierarchical codebook design at the receiver is similar as that at the transmitter, we mainly focus on the design of $\boldsymbol{V}_{\rm t}$. Denote the $i (i=1,2,\cdots,N_{\rm t})$th codeword at the bottom layer of a binary codebook $\boldsymbol{V}_{\rm t}$ as
\begin{equation}\label{f_i}
\setlength{\abovedisplayskip}{3pt} 
\setlength{\belowdisplayskip}{3pt}
\boldsymbol{f}_i=\alpha(N_{\rm t},-1+(2i-1)/N_{\rm t})
\end{equation}
which is essentially a channel steering vector with beam coverage of $[-1+2(i-1)/N_{\rm t},-1+2i/N_{\rm t}]$~\cite{Xiao2016Hierarchical}. In fact, beam coverage is essentially the range of the main lobe covers. As illustrated in Fig.~\ref{fig:HierarchicalCodebook}, the coverage of codewords at the upper layer of $\boldsymbol{V}_{\rm t}$ is a union of the coverage of several codewords at the bottom layer. 
To design a normalized codeword $\boldsymbol{v}=\boldsymbol{V}_{\rm t}(s,m)$, the set of the indices of $\boldsymbol{f}_i$ for the union can be determined as
\begin{equation}\label{indices of f_i}
\setlength{\abovedisplayskip}{3pt} 
\setlength{\belowdisplayskip}{3pt}
\boldsymbol{\Psi}_{s,m}=\bigg\{i\bigg|\frac{(m-1)N_{\rm t}}{2^{s}}+1\le i\le \frac{mN_{\rm t}}{2^{s}},~i=1,2,\cdots N_{\rm t} \bigg\}.
\end{equation}
Then we can figure out a general codeword as a weighted summation of channel steering vectors as
\begin{equation}\label{designed codeword without normalization}
\setlength{\abovedisplayskip}{3pt} 
\setlength{\belowdisplayskip}{3pt}
\boldsymbol{\hat{v}}=\sum_{i\in\boldsymbol{\Psi}_{s,m}}e^{j\theta_i}\boldsymbol{f}_i
\end{equation}
where $\theta_i$ is used to avoid low beam gain within the beam coverage~\cite{PS-DFT2017}.
Given $\boldsymbol{\hat{v}}$, $\boldsymbol{v}$ can be obtained by
\begin{equation}\label{designed codeword with normalization}
\setlength{\abovedisplayskip}{3pt} 
\setlength{\belowdisplayskip}{3pt}
\boldsymbol{{v}}=\frac{\boldsymbol{\hat{v}}}{\|\boldsymbol{\hat{v}}\|_2}.
\end{equation}
regarding the fact that the codewords do not provide power gain. The beam gain of $\boldsymbol{\hat{v}}\in\mathbb{C}^{N_{\rm t}}$ along $\Omega$ is typically written as
\begin{equation}\label{beam gain}
\setlength{\abovedisplayskip}{3pt} 
\setlength{\belowdisplayskip}{3pt}
G(\boldsymbol{\hat{v}},\Omega)=\sqrt{N_{\rm t}}\alpha(N_{\rm t},\Omega)^H\boldsymbol{\hat{v}}=\sum_{n=1}^{N_{\rm t}}[\boldsymbol{\hat{v}}]_ne^{-j\pi(n-1)\Omega}.
\end{equation}

\begin{figure}[!t]
  \centering
  \includegraphics[width=90mm]{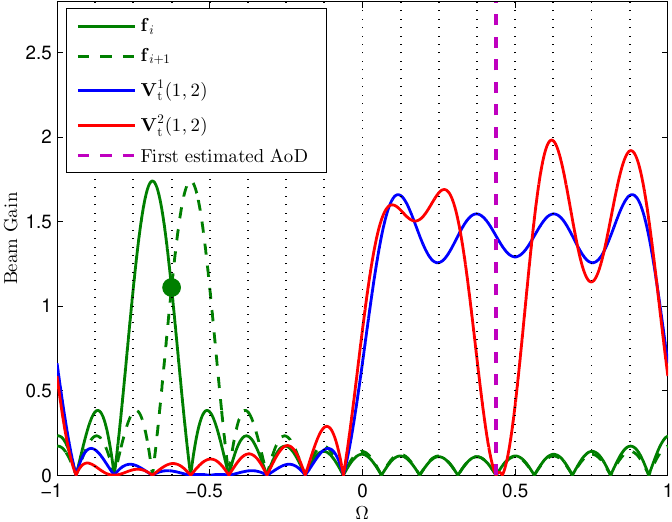}
  \caption{Beam gain of designed codewords from generalized hierarchical codebook.}
  \label{fig:CodewordDesign}
\end{figure}
Without loss of generality, we pick up two channel steering vectors from $\boldsymbol{\Psi}_{s,m}$, i.e., $\boldsymbol{f}_i$ and $\boldsymbol{f}_{i+1}$ with the directions along $-1+(2i-1)/N_{\rm t}$ and $-1+(2i+1)/N_{\rm t}$, respectively. Therefore the absolution beam gain along $-1+(2i-1)/N_{\rm t}$ and $-1+(2i+1)/N_{\rm t}$ is high, which is illustrated in Fig.~\ref{fig:CodewordDesign}. To avoid the low beam gain within the beam coverage, the beam gain at the middle angle of $\boldsymbol{f}_i$ and $\boldsymbol{f}_{i+1}$, i.e., $-1+2i/N_{\rm t}$ that is marked as a green point in Fig.~\ref{fig:CodewordDesign}, is expected to be also high. In this context, we resort to $\theta_i$ and $\theta_{i+1}$. To maximize the beam gain along $-1+2i/N_{\rm t}$, the optimization problem can be formulated as $\underset{\theta_i,\theta_{i+1}}{\max} |e^{j\theta_i}G(\boldsymbol{f}_i,-1+2i/N_{\rm t}) + e^{j\theta_{i+1}}G(\boldsymbol{f}_{i+1},-1+2i/N_{\rm t})|$. Note that $|e^{j\theta_i}G(\boldsymbol{f}_i,-1+2i/N_{\rm t})|=|e^{j\theta_{i+1}}G(\boldsymbol{f}_{i+1},-1+2i/N_{\rm t})|$. We can solve the optimization problem by
\begin{equation}\label{design of theta_i}
\setlength{\abovedisplayskip}{3pt} 
\setlength{\belowdisplayskip}{3pt}
e^{j\theta_i}G\bigg(\boldsymbol{f}_i,-1+\frac{2i}{N_{\rm t}}\bigg)=e^{j\theta_{i+1}}G\bigg(\boldsymbol{f}_{i+1},-1+\frac{2i}{N_{\rm t}}\bigg).
\end{equation}
Then we can obtain $\theta_{i+1}-\theta_i=(-1+1/N_{\rm t})\pi+2k\pi, k\in\mathbb{Z}.$ Without loss of generality, we set $\theta_{i+1}-\theta_i=(-1+1/N_{\rm t})\pi$. Then, the designed codeword in \eqref{designed codeword without normalization} can be rewritten as
\begin{equation}\label{designed codeword without normalization2}
\setlength{\abovedisplayskip}{3pt} 
\setlength{\belowdisplayskip}{3pt}
\boldsymbol{\hat{v}}=\sum_{i\in\boldsymbol{\Psi}_{s,m}}e^{j\pi(-1+{1}/{N_{\rm t}})i}\boldsymbol{f}_i.
\end{equation}
In \eqref{designed codeword without normalization2}, we set $\theta_{1}=(-1+1/N_{\rm t})\pi$ for simplicity, which does not change the absolute beam gain of $\boldsymbol{\hat{v}}$. Based on $\boldsymbol{\hat{v}}$, $\boldsymbol{{v}}$ can be obtained via \eqref{designed codeword with normalization}.

Although we only pick up two channel steering vectors from $\boldsymbol{\Psi}_{s,m}$, they contribute much more energy than the other neighbouring channel steering vectors, regarding the fact that the energy of the other neighbouring channel steering vectors comes from their sidelobes. In this way, we can simplify the codeword design.
Note that the above method can be used to design codewords at any layer of the hierarchical codebook, leading to generalized hierarchical codebook design. In particular, we may change $\boldsymbol{\Psi}_{s,m}$ to achieve the beam coverage with some points of zero beam gain, which will be utilized in the following subsection.



\begin{algorithm}[!t]
	\caption{Beam Training based on Dynamic Hierarchical Codebook}
	\label{ALG：hierarchical search}
	\begin{algorithmic}[1]
        \STATE \textbf{Input:} $N_{\rm t}$, $N_{\rm r}$, $L_{\rm d}$.
        \STATE \textbf{Initialize:} $\boldsymbol{T} \leftarrow \varnothing$, $\boldsymbol{R} \leftarrow \varnothing$.
        \STATE Obtain $\boldsymbol{\Psi}^1_{s,m}$ and $\boldsymbol{\Phi}^1_{s,m}$ via \eqref{indices of f_i}.
        \STATE Obtain $\boldsymbol{V}_{\rm t}^1(s,m)$ and $\boldsymbol{V}_{\rm r}^1(s,m)$ via \eqref{designed codeword without normalization2} and \eqref{designed codeword with normalization}.
        \FOR{$l=1,2,\ldots,L_{\rm d}$}
        \STATE  Perform beam training with $\boldsymbol{V}_{\rm t}^l(s,m)$ and $\boldsymbol{V}_{\rm r}^l(s,m)$.
        \STATE Update $\boldsymbol{T}$ and $\boldsymbol{R}$ via \eqref{recordMPCs}.
        \STATE Obtain $\boldsymbol{\Psi}_{s,m}^{l+1}$ and $\boldsymbol{\Phi}_{s,m}^{l+1}$ via \eqref{set at the transimitter and receiver}.
        \STATE Obtain $\boldsymbol{V}_{\rm t}^{l+1}(s,m)$ and $\boldsymbol{V}_{\rm r}^{l+1}(s,m)$ via \eqref{designed codeword without normalization2} and \eqref{designed codeword without normalization}.
        \ENDFOR
        \STATE \textbf{Output:} $\boldsymbol{T},\boldsymbol{R}$.
	\end{algorithmic}
\end{algorithm}

\subsection{Beam Training based on Dynamic Hierarchical Codebook}
For a multi-path channel with $L$ MPCs, existing beam training methods normally use the same hierarchical codebook for the estimation of different MPCs, while subtracting the ``estimated'' part before the new estimation according to \eqref{system model3}. However, it relies on the precision of the ``estimated'' part. In this work, we use different hierarchical codebooks instead of the same hierarchical codebook for the estimation of different MPCs. Before estimating a new MPC, the codebook is dynamically updated according to the estimated MPCs. By introducing the dynamic hierarchical codebook, the operation of substraction can be omitted.

Suppose $L_{\rm d}$ MPCs are expected to be estimated after the beam training. Note that $L_{\rm d}$ is usually determined by the prespecified number of data streams that are transmitted in parallel during mmWave massive MIMO communications~\cite{heath2016overview}. For the $l(l=1,2,\cdots,L_{\rm d})$th MPC, the hierarchical codebooks employed at the transmitter and receiver are denoted as $\boldsymbol{V}_{\rm t}^l$ and $\boldsymbol{V}_{\rm r}^l$, respectively. The sets of indices of $\boldsymbol{f}_i$ for the union in $\boldsymbol{V}_{\rm t}^l(s,m)$ and $\boldsymbol{V}_{\rm r}^l(s,m)$ are denoted as $\boldsymbol{\Psi}^l_{s,m}$ and $\boldsymbol{\Phi}^l_{s,m}$, respectively. Using the existing beam training method based on hierarchical codebook $\boldsymbol{V}_{\rm t}^l$ and $\boldsymbol{V}_{\rm r}^l$~\cite{Xiao2016Hierarchical,ZhenyuXiao2018}, the AoA and AoD for the $l$th MPC can be estimated.

 We start the first round beam training with two initial codebooks denoted as $\boldsymbol{V}_{\rm t}^1(s,m)$ and $\boldsymbol{V}_{\rm r}^1(s,m)$, which can be designed once $\boldsymbol{\Psi}^1_{s,m}$ and $\boldsymbol{\Phi}^1_{s,m}$ are obtained via \eqref{indices of f_i}.  After the $l$th beam training, we suppose that the estimated channel AoA and AoD correspond to the $q$th and $p$th channel steering vectors, which are the information of the $l$th MPC. To record the estimated MPCs, we introduce two sets $\boldsymbol{T}$ and $\boldsymbol{R}$ at the transmitter and receiver, respectively. We iteratively update $\boldsymbol{T}$ and $\boldsymbol{R}$ as
\begin{equation}\label{recordMPCs}
\setlength{\abovedisplayskip}{3pt} 
\setlength{\belowdisplayskip}{3pt}
  \boldsymbol{T}=\boldsymbol{T} \cup p,~~\boldsymbol{R}=\boldsymbol{R} \cup q
\end{equation}
where both $\boldsymbol{T}$ and $\boldsymbol{R}$ are initialized to be empty sets.

Then we  dynamically update the  hierarchical codebook for the $(l+1)$th beam training as
\begin{align}\label{set at the transimitter and receiver}
\setlength{\abovedisplayskip}{3pt} 
\setlength{\belowdisplayskip}{3pt}
\boldsymbol{\Psi}_{s,m}^{l+1}=\boldsymbol{\Psi}_{s,m}^l \backslash p, ~~
\boldsymbol{\Phi}_{s,m}^{l+1}=\boldsymbol{\Phi}_{s,m}^l \backslash q
\end{align}
In \eqref{set at the transimitter and receiver}, we remove the $l$th estimated MPC from the hierarchical codebooks used for the $l$th beam training. The motivation is to remove the contribution of the estimated MPCs before we start to estimate a new MPC. In this way, we can skip the estimation of the channel gain $\lambda_l$ of the estimated MPCs, which are required in the substraction of the ``estimated'' part from the received signal according to~\eqref{system model3}.

As shown in Fig.~\ref{fig:CodewordDesign}, we illustrate $\boldsymbol{V}^1_{\rm t}(1,2)$ and $\boldsymbol{V}^2_{\rm t}(1,2)$ with $N_{\rm t}=16$.  We first obtain $\boldsymbol{\Psi}_{1,2}^1=\{8,9,\ldots,16\}$ via \eqref{indices of f_i}. Then $\boldsymbol{V}^1_{\rm t}(1,2)$ can be designed via \eqref{designed codeword without normalization2} and is marked by a blue solid line. Suppose the estimated AoD after the first round beam training corresponds to the $12$th codeword at the bottom layer of $\boldsymbol{V}_{\rm t}^1$. To remove the contribution of the first estimated MPC at the transmitter before starting to estimate the second MPC, we delete 12 from $\boldsymbol{\Psi}_{1,2}^1$ according to \eqref{set at the transimitter and receiver} and obtain $\boldsymbol{\Psi}_{1,2}^2=\{8,9,10,11,13,14,15,16\}$. The beam gain of the resulting $\boldsymbol{V}^2_{\rm t}(1,2)$ is marked by a red solid line. It is seen that compared to the beam gain of $\boldsymbol{V}^1_{\rm t}(1,2)$, the beam gain of $\boldsymbol{V}^2_{\rm t}(1,2)$ is set zero along the angle of $7/16$.

\begin{figure}[!t]
  \centering
  \includegraphics[width=90mm]{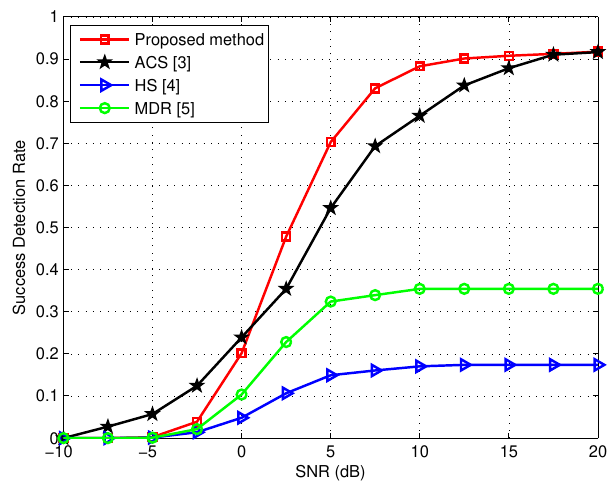}
  \caption{Comparisons of success detection rate for different beam training methods.}
  \label{fig:SuccessDetectionRate}
\end{figure}

The procedures of beam training based on dynamic hierarchical codebook are summarized in \textbf{Algorithm~\ref{ALG：hierarchical search}}.
Suppose the beam training is started from $S_0$th layer. We need to update $(\log_2{N_{\rm t}}-S_0+1)$ codewords for the $l(l\ge2)$th MPC. For example, in our simulation with $N_t=32$ and $S_0=2$, we only need to update $(\log_2{N_{\rm t}}-S_0+1)=4$ codewords, which does not require a large amount of computation. In fact, we only need to update the codewords used for the next beam training instead of updating all involved codewords in the codebook, which can further reduce the computational complexity of updating the codebooks.

\section{Simulation Results}
We consider an mmWave massive MIMO system equipped with the same number of transmitting and receiving antennas, i.e., $N_{\rm t}=N_{\rm r}=32$. The mmWave massive MIMO channel is setup with $L=3$ non-line-of-sight (NLOS) channel paths, where $\lambda_l\sim\mathcal{CN}(0,1)$ for $l=1,2,3$~\cite{ZhenyuXiao2018}. Both the channel AoD $\Omega_l^{\rm t}$ and the channel AOA $\Omega_l^{\rm r}$ obey the uniform distribution within $[-1,1]$. For the proposed beam training method, we set $L_{\rm d}=3$ and $S_0=2$.

As shown in Fig.~\ref{fig:SuccessDetectionRate}, we compare the proposed method with the existing ACS~\cite{Sparse2014}, HS~\cite{Xiao2016Hierarchical} and MDR~\cite{ZhenyuXiao2018}, in terms of success detection rate of beam training. The success detection rate of beam training is defined as follows. If all the channel AoA and AoD of $L_{\rm d}$ MPCs are correctly identified, we say the beam training is successful; otherwise, it is failed. It is seen from Fig.~\ref{fig:SuccessDetectionRate} that the proposed method performs much better than HS and MDR.
Although the performance of the proposed method is similar as ACS in high SNR region, e.g., SNR $\rm=20$ dB, the former outperforms the latter in the low SNR region, e.g., the former can achieve almost $20\%$ improvement over the latter at SNR $\rm=7.5$~dB.

Aside of performance comparisons, we also compare their overhead of beam training. The total number of beam training for the proposed method is $L_{\rm d}(4^{S_0}+4\log_2{N_{\rm t}}-4S_0)$, which is the same as HS. The total number of beam training for MDR and ACS is $L_{\rm d}(4^{S_0}+4\log_2{N_{\rm t}}-4S_0+9)$ and $K^2L_{\rm d}^3\log_2 N_{\rm t}$, respectively. In our simulations with $L_{\rm d}=3$, $N_t=32$, $K=2$ and $S_0=2$, the total number of beam training of the proposed method, HS, MDR and ACS is $84$, $84$, $111$ and $540$, respectively. Therefore, using the proposed method can reduce the overhead of beam training by $24\%$ and $84\%$ compared to MDR and ACS, respectively. The reason is that to obtain an estimate of the channel gain, MDR needs additional beam training while ACS needs higher resolution of codebook. However, the proposed method avoids the estimation of the channel gain, leading to the reduction of overhead of beam training compared to MDR and ACS.

\section{Conclusion}
In this letter, we have proposed a beam training method based on dynamic hierarchical codebook for mmWave massive MIMO. A generalized hierarchical codebook design method has been proposed. Then based on this method, we have proposed a beam training method which dynamically updates the hierarchical codebook by removing the contribution of the estimated MPCs from the codebook. Future work will be continued with the focus on parallel data transmission after the beam training.

\bibliographystyle{IEEEtran}
\bibliography{IEEEabrv,IEEEexample}

\begin{thebibliography}{1}
\providecommand{\url}[1]{#1}
\csname url@samestyle\endcsname
\providecommand{\newblock}{\relax}
\providecommand{\bibinfo}[2]{#2}
\providecommand{\BIBentrySTDinterwordspacing}{\spaceskip=0pt\relax}
\providecommand{\BIBentryALTinterwordstretchfactor}{4}
\providecommand{\BIBentryALTinterwordspacing}{\spaceskip=\fontdimen2\font plus
\BIBentryALTinterwordstretchfactor\fontdimen3\font minus
  \fontdimen4\font\relax}
\providecommand{\BIBforeignlanguage}[2]{{%
\expandafter\ifx\csname l@#1\endcsname\relax
\typeout{** WARNING: IEEEtran.bst: No hyphenation pattern has been}%
\typeout{** loaded for the language `#1'. Using the pattern for}%
\typeout{** the default language instead.}%
\else
\language=\csname l@#1\endcsname
\fi
#2}}
\providecommand{\BIBdecl}{\relax}
\BIBdecl

\bibitem{CL1}
J.~Zhu, P.~Yang, Y.~Xiao, X.~Lei, and Q.~Chen, ``Low {RF}-complexity receive
  spatial modulation for millimeter-wave {MIMO} communications,'' \emph{{IEEE}
  Commun. Lett.}, vol.~22, no.~7, pp. 1338--1341, Jul. 2018.

\bibitem{ShuFengTWC}
F.~Shu, Y.~Qin, T.~Liu \emph{et~al.}, ``Low-complexity and high-resolution
  {DOA} estimation for hybrid analog and digital massive{ MIMO} receive
  array,'' \emph{{IEEE} Trans. Commun.}, vol.~66, no.~6, pp. 2487--2501, Jun.
  2018.

\bibitem{Sparse2014}
A.~Alkhateeb, O.~E. Ayach, G.~Leus, and R.~W. Heath, ``Channel estimation and
  hybrid precoding for millimeter wave cellular systems,'' \emph{{IEEE} J. Sel.
  Top. Signal Process.}, vol.~8, no.~5, pp. 831--846, Oct. 2014.

\bibitem{Xiao2016Hierarchical}
Z.~Xiao, T.~He, P.~Xia, and X.~G. Xia, ``Hierarchical codebook design for
  beamforming training in millimeter-wave communication,'' \emph{{IEEE} Trans.
  Wireless Commun.}, vol.~15, no.~5, pp. 3380--3392, May. 2016.

\bibitem{ZhenyuXiao2018}
Z.~Xiao, H.~Dong, L.~Bai, P.~Xia, and X.~Xia, ``Enhanced channel estimation and
  codebook design for millimeter-wave communication,'' \emph{{IEEE} Trans. Veh.
  Technol.}, vol.~67, no.~10, pp. 9393--9405, Oct. 2018.

\bibitem{heath2016overview}
R.~W. Heath, N.~Gonzalez-Prelcic, S.~Rangan, W.~Roh, and A.~Sayeed, ``An
  overview of signal processing techniques for millimeter wave {MIMO}
  systems,'' \emph{{IEEE} J. Sel. Top. Signal Process.}, vol.~10, no.~3, pp.
  436--453, Apr. 2016.

\bibitem{PS-DFT2017}
S.~Noh, M.~D. Zoltowski, and D.~J. Love, ``Multi-resolution codebook and
  adaptive beamforming sequence design for millimeter wave beam alignment,''
  \emph{{IEEE} Trans. Wireless Commun.}, vol.~16, no.~9, pp. 5689--5701, Sep.
  2017.

\end{thebibliography}
\end{document}